\documentstyle[aasms4,12pt,epsf]{article} 
		 	
\begin{document}  
\tighten
\eqsecnum
\received{}  
\accepted{}  
\journalid{}{}
\articleid{}{}	
\def\msol{$\rm M_\odot$}  
\def\ltsima{$\; \buildrel < \over \sim \;$}  
\def\simlt{\lower.5ex\hbox{\ltsima}}             
\def\gtsima{$\; \buildrel > \over \sim \;$}  
\def\simgt{\lower.5ex\hbox{\gtsima}}      
\def\kms{km~s$^{-1}$}  
\def\ergs{ergs~s$^{-1}$\/}
\def\a{$\alpha$}  
\def\b{$\beta$}  
\def\l{$\lambda$}  
\def\ha{{\sc H}$\alpha$}
\def\hb{{\sc{H}}$\beta$\/} 
\def\hbbc{{\sc{H}}$\beta_{\rm BC}$\/}   
\def\hbnc{{\sc{H}}$\beta_{\rm NC}$\/} 
\def\civ{{\sc{Civ}}$\lambda$1549\/}  
\def\feiiuv{{{\sc{Feii}}}$_{\rm UV}$\/} 
\def\feii{{\sc{Feii}}$_{\rm opt}$\/} 
\def\heiiuv{{\sc{Heii}}$\lambda$1640} 
\def\heiiuvnc{{\sc{Heii}}$\lambda$1640$_{\rm NC}$\/} 
\def\heii{{\sc{Heii}}$\lambda$4686\/}   
\def\mgii{{\sc{Mgii}}$\lambda$2800\/} 
\def\niv{{\sc{Niv]}}$\lambda$1486\/} 
\def\oiiiopt{{\sc{[Oiii]}}$\lambda\lambda$4959,5007\/}
\def\oiii{{\sc [Oiii]}$\lambda$5007}
\def\OIIIgen{{\sc [Oiii]}}
\def\oiii49{{\sc{[Oiii]}}$\lambda$4959\/}
\title{\sc Balmer Line Variations in the Radio-Loud AGN PG~1512+370}
\author{P. Romano\altaffilmark{1,}\altaffilmark{2}, 
P. Marziani\altaffilmark{3}, D. Dultzin-Hacyan\altaffilmark{4}}
\altaffiltext{1}{Department of 
Physics \&\ Astronomy, Gallalee Hall, Tuscaloosa, AL 35487} 
\altaffiltext{2}{Department of Astronomy, Ohio State University, 
174 West 18th Avenue, Columbus, OH 43210 \\
Electronic mail: promano@astronomy.ohio-state.edu}
 \altaffiltext{3}{Osservatorio Astronomico di Padova, 
Vicolo dell' Osservatorio 5, I-35122 Padova, Italia \\ 
Electronic mail: marziani@astrpd.pd.astro.it}
\author{~ }   
\altaffiltext{4}{Instituto de Astronomia, UNAM, 
Apt.do Postal 70-264, Mexico, DF 04510,
Mexico \\ 
Electronic mail: deborah@astroscu.unam.mx}

\begin{abstract} 

We present  spectroscopic observations of the quasar PG~1512+370,  covering
the  \hb\ line spectral range and collected at moderate  resolution (2-7
\AA\ FWHM) from 1988 to 1996.  The  observations show that the  blue wing of
the \hb\ broad  profile component has changed significantly in flux 
and shape between 1988 and 1990 and between 1995 and 1996.  A displaced blue peak  
on the \hb\ profile, visible in 1988, but not in the 1990-1995 spectra, is
revealed  again in one of the spectra obtained in 1996. The blue peak
(in both the 1988 and 1996 spectra) is centered at $\rm \Delta v_r \approx
-3000^{+500}_{-500}$ \kms\ from the  rest frame defined by the narrow component of \hb,
and the {\sc{[O\,iii]}}\,$\lambda\lambda$4959,5007 lines. 

We discuss several conflicting  interpretations of the data.  
We find  that   
the variability of the \hb\ blue wing is consistent with Balmer line emission
from regions whose motion is predominantly radial, if variations of the blue
wing are a response to continuum changes. Alternatively, we note that observed
\hb\ line profile variations  are consistent with a variable line component  as
in a ``binary black hole'' scenario.  More frequent observations of \hb\ are
needed to distinguish among these hypotheses.

\end{abstract}  


\section{Introduction}   

Understanding the structure of the inner parsec of  quasars--which includes 
the line and continuum emitting regions--has proven elusive.
Reverberation mapping techniques have been applied to several low-luminosity
radio-quiet AGN. Though they required large amounts of telescope time, they set on a firm
ground basic aspects of the Broad Line Region (BLR), i.e., the
stratification of ionization (see \cite{NP97}, \cite{Peterson93}, and \cite{Penston91} for
reviews). Other results are unfortunately more ambiguous. 
One  reason for the ambiguity resides  in our poor understanding of 
the constraints on the inputs to the line
formation processes in the  BLR  (\cite{Baldwin97}).  
Statistical studies of line profiles have in principle  the
advantage of an approach less  dependent on the poorly known physical
conditions of the BLR. However they are as yet plagued by small number statistics and by
the difficulty of obtaining a uniform sample, in terms of observational (i.e., 
signal-to-noise ratio, S/N) and intrinsic properties (such as optical luminosity; 
see \cite{Baldwin97} for a recent review).

Radio-loud  AGN have been
associated to an extreme in the distribution of parameters of radio-quiet AGN
(\cite{Brotherton96}; \cite{CorbinBoroson96}). However, Marziani et al. (1996,
hereafter M96) have shown that radio-loud AGN with detected superluminal motion,
and apparent velocity $\beta_{app} \simgt 10$\ show \civ\ profiles with low EW
and remarkably strong red-ward asymmetry. There is no equivalent phenomenology
in radio-quiet objects. Also, widely spaced double-peaked profiles appear
almost exclusively in radio-loud AGN (see \cite{Eracleous97a} for a critical review, 
and \cite{EH94}). 
Statistical analysis of Balmer  line profiles   suggests systemic,
kinematic, and structural differences in the BLR of (at least some) 
radio-quiet and radio-loud AGN (M96; \cite{Romano96}).

There is therefore little evidence that the basic reverberation and profile 
analysis results can be extended straightforwardly to radio-loud or high luminosity AGN.
In the handful of radio-loud AGN known to have variable profiles, the  profiles
are often peculiar (i.e., double-peaked), as in the cases of 3C 390.3 
(\cite{Veilleux91}), her twin 3C 382, OQ 208 (\cite{Marziani93}), and Pictor A
(\cite{Sulentic95}; \cite{HE94}). This is different from the case for
radio-quiet AGN, since monitored Seyfert nuclei do not show  
double-peaked line profiles. 
The most common  profile among radio-loud AGN, however,  is not the extremely
broad, double-peaked profile observed in Arp 102B; instead, it is regular,
single peaked, red-ward asymmetric with red-ward peak displacement (M96; class
AR,R according to \cite{Sulentic89}). The double-peaked profiles are probably not
representative of the radio-loud AGN as a class. 

The low redshift (heliocentric $\rm z = 0.3707 \pm 0.0002$; M96) radio quasar 
PG~1512+370 (B2~1512+37, 4C~37.43) has
been frequently observed at optical wavelengths (\cite{OBG91}; 
\cite{BG92}, among others). The \hb\ broad  profile component (\hbbc) 
 of PG~1512+370 belongs
to the class AR,R for most epochs of observations, and hence it is fairly
typical of radio-loud AGN. Variability of broad UV lines \civ\ and Ly$\alpha$\
has been reported in 1991 from IUE observations (\cite{OBG91}).
These authors suggested the possibility of moderate \hb\ line flux variations
from  published data (EW(\hbbc)  appeared to vary by $\approx$ 20 \%). 

In this paper, we report the discovery  of  \hb\ line profile variations in
PG~1512+370, and in particular variation of the \hbbc\ blue wing and of a
secondary ``blue peak'' from the analysis of spectra collected from 1988
to 1996 (\S \ref{results}).  We will attempt to explain the \hbbc\
profile and its variations as due to (a) accretion disk emission (\S
\ref{disk}); (b) radially moving gas (\S \ref{reverber}),  or  
(c) a binary BLR (\S  \ref{bh}). 

\section{Data and Results \label{obs} \label{results}}

We summarize the available data in Table~\ref{tab:observations}.
The initial reduction followed the standard {\tt IRAF}
procedures, which include bias subtraction, flat fielding, wavelength
calibration,  correction for extinction and flux calibration (also see M96, 
and references therein for further details).  The spectra have been corrected
for redshift, the individual redshifts being determined as the average of the
peaks of the  Gaussian fits to the upper half of {\sc{H}}$\beta$\ and
{\sc{[O\,iii]}}\,$\lambda\lambda$4959,5007. 

\begin{figure}
	\null
	\epsfxsize=15truecm
	\epsfysize=15truecm
	\hspace{+0.5truecm}
 	\epsfbox{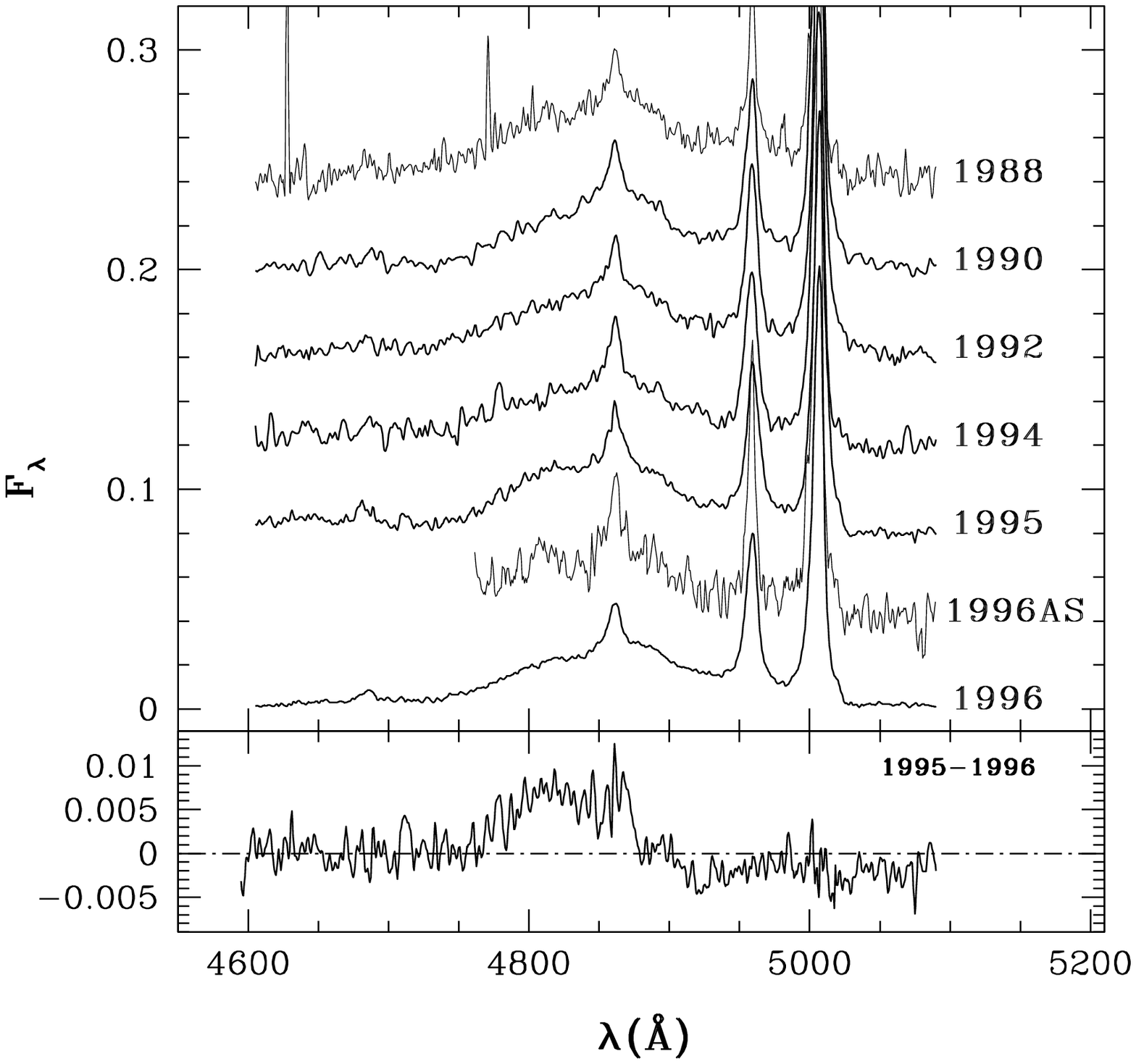}
\caption{Upper Panel: optical spectra of PG~1512+370 
obtained during 1988, 1990, 1992, 1994, 1995 
and 1996 (see Table~\ref{tab:observations} for the observation log). 
Specific flux $\rm F_\lambda$\ is  in  units
of $6.9\times 10^{-15}$ erg~s$^{-1}$~cm$^{-2}$~\AA$^{-1}$ (i.e., the average
flux of {\sc [Oiii]}$\lambda$4959 used as a scaling factor). 
The spectra have been vertically 
offset for clarity, with the exception of the spectrum labeled ``1996''.
Lower Panel: difference between the 1995 and 1996 spectra. Units are the same
as for the upper panel.
\label{fig:spectra}} 
\end{figure}

The upper panel of Figure~\ref{fig:spectra} shows the optical spectra for
PG~1512+370 in the {\sc{H}}$\beta$\  region. The spectra have been continuum
subtracted by fitting a low order  polynomial through wavelength regions
uncontaminated by emission lines. Most of the weight in the continuum fit
has been assigned  to: (1) the region  around 4200 \AA{} as adopted for the
quasar composite spectrum by Francis et al. (1991), and (2) to the red side of
{\sc{[O\,iii]}}\,$\lambda$5007.  Other wavelength regions around 4020 \AA{}
and, with decreasing weight,  around  4620 \AA{} have been included in the
fit when available.  The spectra have then been scaled   to the same flux of
{\sc{[O\,iii]}}\,$\lambda$4959.  
We did not choose
{\sc{[O\,iii]}}\,$\lambda$5007 because of the intrinsically larger errors
due to the presence of the telluric B band at its red side. The scale factor
determined on the basis of  {\sc{[O\,iii]}}\,$\lambda$4959   is 
affected by the extended  appearance of the {\sc{[O\,iii]}} emission
(diameter $\approx$ 4 arcsecs, \cite{Durret94}). On the other hand, the 
\hb\ narrow profile component (\hbnc) 
flux is difficult to measure, because the underlying \hbbc\ is
variable. As a result, measured fluxes  are thought to be accurate within
$\pm 10$\%\ at a $3 \sigma$ confidence level.  The  procedure to remove
the contribution of 
\feii{} (which is minimal; see, e.g.,  \cite{BG92}), \heii,  
\hbnc\ and \oiiiopt\ from the \hb\ profile is
described in M96. Table~\ref{tab:lines} reports measurements taken on the 
cleaned \hbbc\ line.

The only signicant change of {\em total} \hbbc\ flux recorded in our spectra 
occurred probably between 1988 and 1990, when it was $\approx 30$\%\ of the 1988
flux. The larger decrease is observed in the blue wing of \hbbc\  and with a 
possible smaller
decrease in the \hbbc\ red wing. Between 1990 and 1996 the \hbbc\ flux 
remained  constant within the observational
uncertainties. No  shape change  is noticeable in the 1990, 1992, and 1994
spectra. At these times, there is no  peak  at $\rm \Delta v_r \approx -3000$~\kms\
(the ``blue peak'') or blue shoulder visible and  the
\hbbc\ profile can be classified as AR,R according to Sulentic (1989). However,
the line profile showed remarkable changes between 1988 and 1990 as well as between
1995 and 1996.  

\begin{enumerate}
\item The blue peak is visible in the 1988 spectrum but is apparently absent
in 1990, when the total \hbbc\ flux is also significantly lower.
\item The blue wing appears more prominent in the 1995 spectrum
but depressed in the 1996 spectrum obtained at SPM, with a blue wing fractional flux
decrease of $\approx$ 20\%\ .{} 
This variation resembles the one that occurred between
1988 and 1990, save that the total \hbbc\ flux did not change appreciably in the 1995
and 1996 observations.
\end{enumerate}
The lower panel of Figure~\ref{fig:spectra} shows the difference between 
the 1995 and 1996 spectra in the \hb\ region. 
The slight depression in the 4925-5025 \AA{} region is due to the 
extended nature of \OIIIgen, as different contribution from extra-nuclear 
\OIIIgen{} probably fell within the slit in the 1995 and 1996 spectra.  
To compute the probability that the observed
change in line profile shape is due to random noise fluctuation, we first
estimated the 3$\sigma$\ noise level from the observed maximum noise
fluctuations over the average signal in the rest frame range 5060-5100 \AA\ 
(much larger than the noise correlation length $\rm \lambda_{C,
Noise} \approx 4$ pixels).  We rebinned the spectrum  (over the wavelength range
where we observe a change) with a step  $\approx \rm \lambda_{C,Noise}$. We
then multiplied the probability that  the change in amplitude for each bin is due
to noise (which is assumed constant over the \hbbc\ line wing). We find 
the probability that the blue wing change between 1995-1996 is due
to random noise fluctuations is negligibly small. Similar considerations
apply to the 1990-1988 \hbbc\ line profile difference 
(not shown in Figure~\ref{fig:spectra}).  

The remaining question at this point is whether the blue peak may or may not
be variable in radial velocity. 
The ability to resolve the ``blue shoulder''
into a ``blue peak'' depends on the spectral resolution of the data.   
The  1988 and 1996 (Asiago) spectra have a  resolution of 3 \AA\ and 2 \AA\ FWHM
respectively. The blue peak is well visible at  $\Delta \rm v_r \approx
-2960^{+600}_{-250}$ \kms\ (in 1988)  and $\Delta \rm v_r \approx
-3400^{+750}_{-400}$\ \kms\ (in 1996) with respect to the radial velocity of
\hbnc. The 1995-1996 difference has a broad Gaussian shape, with maximum
at $\lambda \approx 4814$ \AA.  Radial velocity values of the blue peak
measured in 1988, 1995, and 1996 are therefore
consistent. The possibility that the blue peak is present in the 1990, 1992, and
1994 spectra but not visible because of the lower resolution cannot be ruled
out. After all, the blue flat top observed in the 1996 San Pedro Martir 
spectrum is resolved as a
blue peak in the 1996 Asiago spectrum, collected 30 days earlier, when the
blue wing appeared depressed. 
Two interpretations seem possible.
\begin{enumerate}
\item
The blue peak is a  component of \hbbc\ more  variable than the rest of
\hbbc,  at approximately constant $\Delta \rm v_r$. In this case, the blue peak
(i) may be due to a flux redistribution within \hbbc\ unrelated to continuum
changes or  (ii)  may be a structure produced by response to continuum variation.
Both the 1988 and 1995 spectra (when the blue wing was more prominent)
show the highest continuum fluxes. This suggests that an enhancement of the blue
wing is related to an increase of the continuum level. However, as the
total \hbbc\ flux did not change significantly in all our spectra obtained
between 1990 and 1996, it is also possible, albeit less likely, that we are
observing a line flux redistribution unrelated to continuum variations (as in
NGC 5548; \cite{WP96}). Model implications are discussed  in  \S
\ref{reverber}.
\item
If the disappearance of the blue peak in 1990-1994 is
genuine, the radial velocity of the  peak in the blue wing may have changed to $\approx$
0$^{+500}_{-1000}$ \kms\
in 1990, 1992,  and 1994. In \S \ref{bh} we discuss a possible model. 
\end{enumerate}

\section{Discussion \label{disc}} 
\subsection{Accretion Disk Line Emission \label{disk}}

The observed \hbbc\ profile as a whole is not consistent with the predictions 
of disk models at any epoch of observation. 
Nevertheless, if we wanted to interpret the secondary peak
$\Delta \rm v_r \approx -3000$\ \kms{} as the Doppler-boosted peak of a disk
contribution, we  would need a disk with $\rm r_{in} \approx 600, \rm r_{out}
\approx 1000$, $\rm i \approx 21^\circ$\ following Chen \&\ Halpern (1989). 
Non-conventional disk models (elliptical disks or warped disks; 
\cite{Eracleous95}; \cite{Marziani97}) may be more successful in fitting 
the \hb\ line profile at a given epoch. 
However, the
variation observed between 1988 and 1990 and between 1995 and 1996 rules out
the possibility of  ``switched-on'' disk emission, since the \hbbc\ change
appears to be restricted to the blue wing in both cases 
(see Figure~\ref{fig:spectra}).
The observed variation would require implausible changes of the disk 
parameters from one epoch to another.

\subsection{A Radially Moving \hbbc\ Line Component \label{reverber}}

The moderate change in the \hbbc\ blue wing flux, $\approx 20$\%, without
a simultaneous response in the red wing, points toward line emission from
gas whose motion is predominantly radial.   
Given the observed blueshift and assuming optically thin clouds,  
as it is generally believed to be the case for most \hb\ emitting clouds, 
the gas must be moving outwards. We are not saying here that the
\hbbc\ line emission is due exclusively to outflowing gas, but that a
fraction related to the blue peak most probably is.  Without further data,
it is however not possible to elaborate  on this result. Caution is needed
as the temporal sampling of our spectra is quite probably not sufficient to
constrain line and continuum variations (\cite{Peterson93}). The \hbbc\ line
luminosity is L(\hbbc)$ \approx 1.2 \times 10^{43}$ \ergs\ for H$_0 = 75$
\kms\ Mpc$^{-1}$, and q$_0 = 0.1$. Under the assumptions of Menzel-Baker
recombination theory, the light travel time across the BLR is $\rm
\tau_{LT} \approx 107 ~L_{43}(H\beta)^\frac{1}{3} \, n_{11}^{-2}
\,f_{f,-5}^{-1}$\ days. Spectra obtained at a frequency of $\simgt$ 1
month$^{-1}$\ are desirable. If the time scale for continuum variation is
found to be comparable to $\rm \tau_{LT}$, then ``ripples''  in the line
profile could be produced  by reverberation  effects. 

At 20 cm, on a FIRST map (\cite{BWH94}), PG 1512+370 appears as a double-lobed radio
source, with lobes extending up to a projected linear distance of $\sim$ 400 kpc from
the central core, at P. A. $\approx$ 105$^\circ$. Integral field spectroscopy
revealed two \oiiiopt\ emitting regions approximately aligned along the East-West
direction with the nucleus (\cite{Durret94}). Our 1996 spectrum also revealed  the
eastern emitting region  as a ``blob'', at $\Delta \rm v_r \approx -300 $ \kms, in
excellent agreement with the results of Durret et al. (1994). Extended \oiiiopt\
emitting regions in radio quasars show a statistical tendency to align with the
radio-source axis in radio-loud, steep spectrum quasars (\cite{BaumHeckman89}).
Extended \oiiiopt\  provides some evidence in favor of outflowing gas related to the
presence of radio ejecta (see  the discussion by \cite{Durret94} for
PG~1512+370).  We consider it unlikely that the blue peak is due to some out-flowing
gas  that is episodically  related to the radio jet. The dynamical time-scale for the
BLR is probably shorter than 8 years; a blob moving at $\rm v_r \simgt 3000$ \kms\ 
could not have maintained the same radial velocity over that time period.

\subsection{Binary BLR \label{bh}}

Within the errors in our spectra, the \hbbc\ profile variations can be ascribed to a
stationary profile plus a superimposed component of variable radial velocity,
contributing up to $\simlt 10$\%\  of the total \hbbc\ line flux. In a binary BLR
framework (see, e.g., \cite{Gaskell83}, \cite{Gaskell96}, \cite{Gaskell96b}), we 
assume that the
secondary component is observed at $\rm v_r \sim 0$ \kms\ (in any case at  $\rm v_r
\simlt 1000$ \kms) between  1990 and 1994 (i.e., not clearly observed because it
underlies the \hbnc). This assumption is justified since  the \hbbc\ underlying the
\hbnc\ may be variable (Figure~\ref{fig:spectra}), and  a slightly redshifted ``shoulder''
partly underlying \hbnc\ may be identified in the 1996 (SPM) spectrum. The ``shoulder'' 
close to \hbnc\ is less prominent in 1995, when the blue wing was in turn
much stronger. 
This   interpretation is not fully
satisfactory because the 1988 and 1996 Asiago spectra are of higher dispersion than
the remaining spectra.  The blue peak is resolved in the 1996 Asiago spectrum but not in
the 1996 San Pedro Martir spectrum,  because of the lower resolution (the two spectra
were obtained 30 days apart). The same may be true for the 1990, 1992, and 1994
spectra.

If we accept the blue peak as a component of variable radial velocity,  
then the immediate implication 
for the orbit of a binary black hole system is that it is highly eccentric, 
and that the binary must have mass ratio very  different from unity. From
the  observed-frame core-to-lobe flux ratio R = 0.091  measured on the FIRST map 
(1.4 GHz)  
we can estimate the angle between the line of sight and the radio axis.  
It is known that the relationship between R and orientation is valid  in a
statistical sense,  probably because of a  large intrinsic scatter in the  the
Lorentz factor of the jet ($ 2 \simlt \gamma \simlt 20$; \cite{PU92};
\cite{Ghisellini93}), and in the ratio f between the intrinsic jet luminosity and
the unbeamed luminosity (\cite{PU92}).  We can write R as R = f $\rm \delta^p$,
where $\delta$\ is the Doppler factor, and  p = 2.75 for a continuous jet
(\cite{Ghisellini93}). We assume $\gamma = $5, and we use f = $7\times 10^{-3}
$\  (\cite{PU92}), as appropriate for Fanaroff-Riley type II radio sources. We
obtain an inclination $ \rm i \approx 21^\circ$. For more extreme values of
$\gamma$, namely for $\gamma = 10$\ and $\gamma = 2$, we obtain 
i=15$^\circ$~ (f= $7\times 10^{-3}$) and i= 26$^\circ$~ (f=$1\times 10^{-2}$) respectively. 
To assume $ \rm i \approx 21^\circ$\ seems therefore reasonable.
PG 1512+370 may have been excluded  from superluminal motion searches, since its radio
emission is lobe-dominated.  However, at $ \rm i \approx 21^\circ$\ and $\gamma
\approx 5$, we expect that PG 1512+370 should be superluminal with apparent
transverse velocity $\rm \beta_{app} \approx 4$~ (also if $\gamma=2$, which implies
i=26$^\circ$, we should observe $\rm \beta_{app} \approx$ 1.7). 

If $\rm i = 21^\circ$, an acceptable fit to the six observed radial velocities 
($\chi_\nu^2\approx$1.76) is obtained for binary period  $\rm P
= 8.15$ yr, eccentricity $\rm e \approx 0.75$, and semi-major axis  $\rm a = 1.6 
\times10^{16}$\ cm. This solution corresponds to a total mass for the binary of 
$\approx 1.7 \times 10^7$\msol. 
The evolution of massive black hole binaries orbital
parameters (semi-major axis, eccentricity) for several mass ratios has been studied
recently by \cite{Quinlan96}.  From Figure~8 of his work we infer that e $\approx 0.75$
for $\rm a = 1.6 \times10^{16}$\ cm is somewhat too large for any value of the initial
eccentricity of the binary, save possibly for the case of initial eccentricity $ \rm e
= 0.9$. In this case, PG~1512+370 should be in a very-short lived phase of the binary
evolution: high eccentricity drives large gravitational radiation losses that should
lead to the merging of the two black holes in $\sim 10^5$ yr. For the lower power 
Fanaroff-Riley type I radio source  OJ 287 the case for 
a double black hole is stronger (e.g. \cite{Sillanp96}; \cite{Benitez96}), and  rather
similar orbital parameters are found: $e = 0.678 \pm 0.004$, $\rm P = 12.07 \pm 0.01$
yr, $\rm i = 4^\circ\pm2^\circ$\ (\cite{Lehto96}).  If the blazar nature of OJ 287 and
the absence of obvious line emission is due to Doppler-boosted continuum emission,
then  PG 1512+370 and OJ 287 may  both host a binary black hole whose orbital plane is
observed at different inclination.

\section{Conclusion and Open Issues \label{concl}}

We have presented evidence of moderate variations ($\simlt$ 30\% total line flux, with
the largest variations occurring in the blue wing) in the \hbbc\ of PG~1512+370
from the analysis of spectra obtained from 1988 to 1996. A peak displaced by  $\Delta
\rm v_r \approx -3000$ \kms\ with respect to the radial velocity of \hbnc\  (the
``blue peak'') has been
observed in 1988 and  1996.  We have attempted to interpret the observations according
to a number of presently fashionable BLR models:  accretion disk emission (\S
\ref{disk});  radially moving gas, and, in particular, emission due to gas trapped by
the radio jet (\S \ref{reverber});  a binary BLR (\S \ref{bh}).  Only radially moving
gas (but not occasionally associated to the radio jet) and a binary BLR are probably
appropriate. 
This is one of the few cases where evidence from line profiles
may support the presence of a binary BLR (with the cautionary note of \S \ref{bh}), 
although the eccentricity inferred from the observed radial velocity curve may be too
high for the estimated major axis. Like OJ 287, PG 1512+370 would then be a very peculiar
object observed in a very short-lived phase of its existence.  
However, we favor the idea that if \hbbc\ variations are the response to ionizing continuum
changes,  the presence of the blue peak and the observed variations of the \hbbc\ blue
wing may be indicative of radial motion within the BLR.
The blue peak may hence be a ``ripple'' produced by light echoes. 
Indeed, we expect that the blue peak will be detected by future observations 
at the same $\Delta \rm v_r \approx -3000$ \kms\ therefore 
ruling out the  binary black hole model altogether (as in the case of 
3C 390.3; \cite{Eracleous97a}; \cite{Eracleous97b}).
A set of high S/N ratio spectra with a good temporal coverage is necessary 
in order to test the proposed hypotheses.

\acknowledgements  

We thank M. Eracleous, T. Boroson, M. Moles, and P. O'Brien for 
allowing us access to their spectra in the digital form  
and the anonymous referee for suggestions that improved the paper. 
We are also indebted to  J. Sulentic and M. Calvani for suggesting this work 
and for carefully reading the manuscript. 
PR acknowledges financial support from the Universit\`a di Padova and the 
Fondazione Ing.\ Aldo Gini  
during her visit at University of Alabama.
PR also acknowledges the financial support of the Department of
Astronomy at Ohio State University. 
This research has made use of the NASA/IPAC Extragalactic Database (NED) which
is operated by the Jet Propulsion Laboratory, Caltech, under contract  with
the National Aeronautics and Space Administration.

\clearpage			

\clearpage			

\begin{deluxetable}{llcccccccc}
\scriptsize
\tablewidth{0pt}
\tablecaption{Data Sample for PG 1512+370}
\tablehead{
\colhead{Observatory} & \colhead{Dates}  & \colhead{Tel. Ap.}     
& \colhead{Spectr.}  & \colhead{Exp. Time} & \colhead{Slit Width} & 
\colhead{P. A.} & \colhead{Sp. Res.} & 
\colhead{Detector} & \colhead{Ref.} \\
\colhead{} & \colhead{} & \colhead{(m)} & 
\colhead{} & \colhead{(s)} & \colhead{(arcsec)} 
& \colhead{($^{\circ}$)} & \colhead{(\AA)} & 
\colhead{} \\ 
\colhead{(1)} & \colhead{(2)} & \colhead{(3)} 
& \colhead{(4)} & \colhead{(5)} &\colhead{(6)} & 
\colhead{(7)} &\colhead{(8)} & \colhead{(9)} & \colhead{(10)} 
}  
\startdata 
 La Palma   &  1988 May 5  & 2.5  & IDS    & 8000 & 1.6 &  \nodata$^a$ & 3 &  235 camera   & A \nl
 KPNO	    &  1990 Sep 20  & 2.1  & Gold   & 2700 & 1.5 &  0  & 6.5--7 &  TI 800 $\times$ 800  & B \nl
 KPNO	    &  1992 Jul 8   & 4.0  & \nodata$^a$   & 1200 & 1.7 & 90 & 6  &      Tek 1024 $\times$ 1024 & C \nl
 Calar Alto &  1994 Apr 11  & 2.2  & B\&Ch  & 4800 & 2.0 & 90 & 5.6--8 &  Tek 1024 $\times$ 1024 & D \nl
 Calar Alto &  1995 May 1   & 2.2  & B\&Ch  & 5400 & 1.6 & 90 & 6      &  Tek 1024 $\times$ 1024 & E \nl
 Asiago     &  1996 Apr 20  & 1.8 & B\&Ch  & 7200 &  2.0 & 90 &   2   & Thomson 7882 & E \nl 
 San Pedro Martir & 1996 May 20  & 2.2 & B\&Ch & 8400 & 2.1--2.6  & 90 &  6--10 & CCDTEK & E \nl
\tablecomments{$^a$: not recorded.} 
\tablerefs{A: Jackson, Penston \&\ P\'erez 1991;  B: Boroson \&\ Green 1992;  
C: Eracleous \&\ Halpern 1994; D: M96; E: this work.}
\enddata
\label{tab:observations}
\end{deluxetable}

\begin{deluxetable}{lccccc}
\tablewidth{0pt}
\tablenum{2}
\tablecaption{PG 1512+370 H$\beta$\ Line Measurements}
\tablehead{
 \colhead{JD}  & \colhead{EW (H$\beta$)}     
& \colhead{H$\beta$ Flux$^a$}   & \colhead{H$\beta$\ Flux$^a$} & 
\colhead{H$\beta$ Flux$^a$ }  & \colhead{Cont. Specific Flux$^b$}\\
\colhead{} & \colhead{(\AA)} & \colhead{} & 
\colhead{Blue Wing} & \colhead{Red Wing} & \colhead{} \\ 
\colhead{(1)} & \colhead{(2)} & \colhead{(3)} 
& \colhead{(4)} & \colhead{(5)} &\colhead{(6)} }  
\startdata 
2447287  &    94    &      6.72  &   3.49 & 3.21  & 0.73 \\
2448155  &    124   &      5.27 &    2.40& 2.87  & 0.43\\ 
2448812  &    105   &      5.23 &    2.44& 2.79  & 0.49  \\
2449454  &     87   &      4.76 &    2.52& 2.24  & 0.58 \\
2449839  &     65   &      5.18 &   2.87& 2.31 & 0.71 \\
2450194$^c$  &     75$^c$   &      4.65$^c$ &     ... & ...  & 0.49 \\
2450224  &    120   &      5.03 &     2.37& 2.66 &  0.53\\
\tablecomments{$^a$: H$\beta$\ flux in units of [OIII]$\lambda$4959 flux;
$^b$: continuum specific flux 
in units of 10$^{15}$ ergs s$^{-1}$ cm$^{-2}$ \AA$^{-1}$;
$^c$: the H$\beta$\ profile in the Asiago spectrum is truncated at 
$\approx 4750$ \AA. The EW and flux values reported are lower limits.}
\enddata
\label{tab:lines}
\end{deluxetable}

\end{document}